\documentclass[10pt,toc]{iopart}

\usepackage{graphicx}
\usepackage{epsfig}
\usepackage{color}
\usepackage{iopams}
\usepackage{setstack}
\usepackage{subfigure}

\newcommand{\CO}{(Color online) }
\newcommand{\be}{\begin{equation}}
\newcommand{\ee}{\end{equation}}
\newcommand{\bea}{\begin{eqnarray}}
\newcommand{\eea}{\end{eqnarray}}

\newcommand{\Z}{{\mathbb S}}
\newcommand{\Eo}{{{\cal E}_0}}

\newcommand{\eqref}[1]{(\ref{#1})}

\begin{document}

% \begin{frontmatter}

\title[Asymptotic scattering in the one-dimensional three-state quantum Potts model]{Asymptotic scattering and duality in the
  one-dimensional three-state quantum Potts model on a lattice} 

\author{ \'Akos Rapp $^1$, Peter Schmitteckert $^{2,3,8}$, G\'abor Tak\'acs $^{4,5}$, and Gergely Zar\'and $^{6,7,8}$}

\address{$^1$ Institut f. Theoretische Physik, Leibniz Universit\"at Hannover, 30167 Hannover, Germany}
\address{$^2$ Institute of Nanotechnology, Karlsruhe Institute of Technology, 76344 Eggenstein-Leopoldshafen, Germany} 
\address{$^3$ DFG Center for Functional Nanostructures, Karlsruhe Institute of Technology, 76128 Karlsruhe, Germany} 
\address{$^4$ Department of Theoretical Physics, Budapest University of Technology and
Economics, H-1521 Budapest, Hungary}
\address{$^5$ MTA-BME "Momentum" Statistical Field Theory Research Group, H-1521 Budapest, Hungary}
\address{$^6$ Freie Universit\"at Berlin, Fachbereich Physik, Arnimallee 14, D-14195 Berlin, Germany}
\address{$^7$ BME-MTA Exotic Quantum Phases Group, Institute of Physics, Budapest University of Technology and Economics, H-1521 Budapest, Hungary.}
\address{$^8$ Authors to whom correspondence should be addressed.}

\ead{Peter.Schmitteckert@kit.edu,zarand@phy.bme.hu}

% \date{Received 4 October 2012\\
% Published 28 January 2013}

% \journal{ \textit{New Journal of Physics} \textbf{15} (2013) 013058 (25pp)}

\begin{abstract}
We determine numerically  the single-particle and the two-particle
spectrum of the three-state quantum Potts model on a  lattice by using
the density matrix renormalization group method,  and 
extract information on the asymptotic (small momentum) S-matrix of the
quasiparticles.
The low energy part of the finite size spectrum can be 
understood  in terms of a simple effective model introduced in a previous work, 
and is consistent with an asymptotic  S-matrix of an exchange form below
a momentum scale $p^*$.
This scale
appears to vanish faster than the Compton scale, $mc$, as 
one approaches the critical point, suggesting that a 
dangerously irrelevant operator may be responsible for the
 behavior observed on the lattice.
\end{abstract}

% 05.50.+q Lattice theory and statistics (Ising, Potts, etc.) (see also 64.60.Cn Order-disorder transformations, and 75.10.Hk Classical spin models)
% 75.10.Jm Quantized spin models, including quantum spin frustration
% 75.10.Pq	Spin chain models

% \submitto{\NJP}
% \submitto{\JPC ???}
% \submitto{\JPG ???}

\tableofcontents

\section{Introduction}

Being the simplest generalization of the transverse field Ising model,
the $q$ state quantum Potts model is one of the most paradigmatic 
models in statistical physics and quantum field theory. The case of
$q=3$ is somewhat peculiar and is also of particular interest. 
On a regular one dimensional lattice, the $q=3$ state quantum Potts model
displays a second order quantum phase transition between a ferromagnetic state
and a paramagnetic state, just as the transverse field Ising
model~\cite{TFIM1d,PottsRevModPhys,qPotts}. The properties of the critical
state itself are very well characterized: at the critical point, an exact
solution is available~\cite{Baxter:1982}, and the scaling limit is known to be
described in conformal field theory (CFT) by the minimal model of central
charge $C=4/5$~\cite{Belavin:1984vu,Dotsenk:1984,Hamer:1981,Hamer:1988}, 
with the so-called $D_4$ partition function \cite{Cappelli:1986hf}. The ordered and disordered phases
of the quantum Potts model are, on the other hand, much richer than those
of the transverse field Ising model: Similar to e.g. antiferromagnetic chains
of integer spins~\cite{Haldane,SachdevDamleHeis,Sachdevbook}, the gapped
phases (i.e., the ferromagnetic as well as the paramagnetic phase) possess
excitations with internal quantum numbers; as a consequence, the dynamics of
these quasiparticles are much richer than those of the transverse field Ising
model and, in contrast to the critical behavior, 
 the quasiparticle properties of the gapped
phases of the quantum Potts chain are not entirely understood.

\begin{figure}
  \centering
  \includegraphics[width=0.7\textwidth,clip=true]{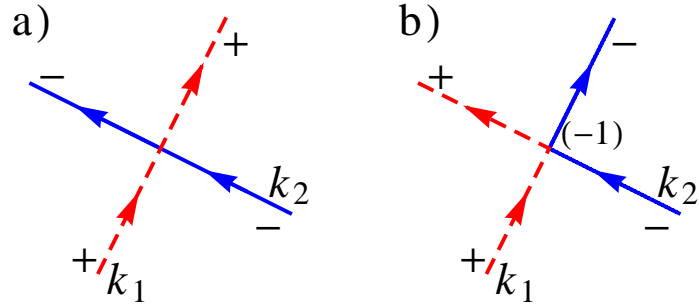}
  \label{collision}
  \caption{\CO Structures of the asymptotic S-matrices. a) Diagonal S-matrix. b) Exchange S-matrix. }
\end{figure}

In the continuum limit, the properties of
the Potts model are usually described by the so-called scaling Potts
field theory, which is a perturbation of the fixed point conformal field
theory, uniquely determined by the symmetries. 
In this approach, the cut-off (lattice spacing) is removed, and
only the leading relevant operator is kept.
The application of  the machinery known as the 
S-matrix bootstrap \cite{ZamolZamol} 
yields a diagonal quasiparticle S-matrix for low energy particles
\cite{Koberle:1979sg,Zamolodchikov:1987zf,PottsTsvelik,Smirnov:1991uw,Pottsintegr,Fendley} 
and implies that the internal quantum numbers of two colliding particles 
are conserved during a scattering process (see  fig.~\ref{collision}.a).
%G i.e., the scattering process should be pure transmission. 
The  bootstrap S-matrix and  the perturbed conformal field theory 
yield a fully consistent picture \cite{Takacs:2011xx}.

%G
%However, recent calculations on the lattice version of the quantum Potts chain
%contradicted these results~\cite{qPotts}: perturbative calculations 
%in both phases
%as well as rather strong renormalization group arguments yielded a
%coherent picture, and supported that in the gapped phases, rather
Recent perturbative calculations as well as renormalization group 
arguments showed~\cite{qPotts}, on the other hand, that rather 
than being diagonal, the asymptotic S-matrix of the {\em lattice  Potts  model}
assumes the "universal" form, also  emerging in various  spin
models~\cite{SachdevDamleHeis}, as well as in the
sine-Gordon model~\cite{ZamolZamol}: $\hat S \to -\hat X$, with  $\hat
 X$ the exchange operator (see fig.~\ref{collision}.b). 
 %G
 %According to this result, quasiparticles
 %of small momenta should scatter on each other by exchanging their
 %quantum numbers (see fig.~\ref{collision}.b). 
 Although the arguments of Ref.~\cite{qPotts} are very robust, the
results of  Ref.~\cite{qPotts} were met by some skepticism. 
On the one hand, the lattice results seemed to  conflict
with  results obtained within the scaling Potts 
theory~\cite{ZamolZamol,Zamolodchikov:1987zf,PottsTsvelik,Smirnov:1991uw,Pottsintegr,Fendley}. 
On the other hand, various thermodynamical properties of the 
two dimensional classical Potts model (on a lattice) such as critical
exponents~\cite{Hamer:1981} or universal amplitude 
ratios~\cite{enting:2003,schur:2002_08}
also seem to agree with the predictions of the scaling Potts 
model (perturbed $C=4/5$ minimal model).
We must emphasize that
 the structure of the asymptotic S-matrix has important physical
 consequences: an S-matrix of the exchange form yields
 diffusive finite temperature spin-spin correlation functions at
 intermediate times~\cite{qPotts,sineGordon,diffusive}, while a
 diagonal S-matrix would result in exponentially damped
 correlations~\cite{Sachdevbook,SachdevYoung,Konik}.

The purpose of the present paper is to investigate and possibly resolve
this apparent controversy. We study in detail the two-particle spectrum of the $q=3$ state
quantum Potts chain using the powerful numerical method of density matrix
renormalization group (DMRG). We find that  the finite size spectra are indeed in complete agreement
with the theory of Ref.~\cite{qPotts} and an asymptotic (i.e., $k\to 0$ momentum) S-matrix of the
exchange form. However, our analysis also reveals the emergence of a new 
momentum scale, $p^*$, below which this exchange S-matrix dominates. 
By approaching the critical point, this scale vanishes faster than the Compton momentum, 
$mc$, suggesting that the new scale and the corresponding exchange scattering is generated by some 
dangerously irrelevant operator, usually neglected in the scaling Potts model.  
Although our numerics are not accurate enough for large momenta,    
they are not inconsistent with a diagonal S-matrix 
as expected within the irrelevant-operator scenario for $|k|\gg p^*$.

\section{The Potts model and its quasiparticles}

In its lattice version, the Potts model consists of a chain of
generalized spins having internal quantum states $|\mu\rangle_i $, with $i$
labeling the lattice sites and $\mu=1,\dots,q$ the possible internal
states of the spins. The Hamiltonian of the $q$-state quantum Potts
chain is then defined as 
\begin{equation}
	H = - J \sum_{i} \sum_{\mu=1}^q P^\mu_i P^\mu_{i+1} - J g
        \sum_i P_i\;. \label{eq:def:H} 
\end{equation}
Here the traceless operators $P_i^\mu =\vert \mu \rangle_{i\,i}\langle \mu
\vert - 1/q$ tend to project the spin at site $i$ along the "direction" $\mu$,
and thus the first term of eq.~(\ref{eq:def:H}) promotes a ferromagnetic
ground state, with all spins spontaneously polarized in one of the directions,
$|\mu\rangle$.  In contrast, the second
term in eq.~(\ref{eq:def:H}) represents a "transverse field", with the
traceless operator $P_i = \vert \lambda_0 \rangle_{i\,i} \langle \lambda_0
\vert-1/q$ trying to align the spins along the direction $|\lambda_0 \rangle
\equiv \sum_\mu \vert \mu \rangle /\sqrt{q}$. The relative strength of these
two terms is regulated by the dimensionless coupling, $g$. These terms
obviously compete with each other, and their competition leads to a phase
transition: for large values of $g$ 
one finds a paramagnetic phase with a
unique ground state, while for small $g$ 
a ferromagnetic phase appears with $q$
degenerate ground states, spontaneously breaking the global $\Z_q$ symmetry. In
the $q=3$ case,--- on which we focus here,--- the transition occurs at a
coupling $g=g_c=1$, and it is of second order: quasiparticles are gapped on
both sides of the transition, but the quasiparticle gap $\Delta$ vanishes
continuously at the transition as $\Delta\sim J\; |g-1|^{5/6}$~\cite{qPotts}.

The $q$ state Potts model obviously possesses a global $\Z_q$
permutation symmetry. As a consequence, the global cyclic permutation
$\mathcal{Z} \vert \mu \rangle_i = \vert \mu+1\;{\rm mod}\;q\rangle_i$
leaves the Hamiltonian also invariant, and can be used to classify its
eigenstates as 
\begin{equation}
	\mathcal{Z} |Q\rangle = e^{i \Omega Q} | Q \rangle , \label{eq:C}
\end{equation}
with $Q$ an integer and the angle $\Omega$ defined as $\Omega =
2\pi/q$. We note that this holds even in the ferromagnetic phase, but there
  states with spontaneously broken symmetries must be mixed.
% \footnote{This holds even in the ferromagnetic phase, but there
%   states with spontaneously broken symmetries must be mixed.} 
In the particular case of $q=3$, considered here, $Q$ can take values of
$Q=0$ and $Q=\pm$. In this case, pairwise spin exchanges (e.g., $\mu =
1\leftrightarrow 2$) also imply that states with quantum numbers
$Q=\pm$ come in degenerate pairs. 

The structure of quasiparticles in the ferromagnetic ($g<1$) and in
the paramagnetic ($g>1$) phases can be easily understood in the
perturbative limits, $g\ll 1$ and $g\gg 1$. For $g>1$ the ground state
$|0)$ is unique, and quasiparticles consist of local spin flips of $\Z_3$
charges $Q=\pm$. For $g<1$, on the other hand, the ground state is
3-fold degenerate,
$|0)\to|0)_\mu$, and quasiparticles correspond to domain walls
between these ground states, $\mu\to \mu' = \mu + \theta\;{\rm
  mod}\;3$, with $\theta=\pm$ the quantum number of the domain wall. 

Similar to the Ising model, the Potts model is known to be
self-dual. 
High-temperature -- low-temperature duality~\cite{dualityC} in the 
$d=2$ classical Potts model implies a duality $g \leftrightarrow 1/g$ 
for the quantum Potts chain~\cite{dualityQ}. In the Appendix we show that
duality holds even on the level of the matrix elements of the
Hamiltonian,  and therefore one can map the spectra 
in the $Q=0$ sectors for $g$ and
$1/g$ by simply rescaling the energies with appropriate factors. We thus have 
\begin{equation}
E_{n}^{Q=0}(g) = g\; E_{n}^{Q=0}(1/g)
\label{eq:duality}
\end{equation}
for all eigenstates $n$ with periodic boundary conditions (PBC), as
also verified later numerically. This duality relation has important
consequences, and shall allow us to relate various energy- and
length scales on the two sides of the transition. 

\section{Effective theory and two-particle S-matrix}

In an infinite system, the elementary excitations of the gapped phases
can be classified by their %momenta
momentum, $k$, and for small momenta their
energy can be approximated as 
\begin{equation}
\epsilon(k) = \Delta + \frac{k^2}{2 m} +\dots\; \label{eq:e(k)}
\end{equation}
independently of their internal quantum number. Here $m=m(g)$ is the
quasiparticle mass, and $\Delta=\Delta(g)$ denotes the quasiparticle
gap. 

In the very dilute limit, interactions between quasiparticles can be
described in terms of just two-body collisions, and correspondingly,
by just two-body scattering matrices and interactions. Assuming
pairwise and short ranged interactions between the quasiparticles, one
thus arrives at the following effective Hamiltonian (in first
quantized form)~\cite{qPotts,Sachdevbook}, 
\begin{equation}
 {\cal H} =  \sum_{i=1}^{N_{qp}} (\Delta - \frac{1}{2m} \frac
 {\partial^2}{\partial  x_i^2})  
+ \sum_{i<j} {u}^{\sigma_i', \sigma_j'}_{\sigma_i,\sigma_j} (x_i -x_j) +\dots\;, \label{eq:Heff}
\end{equation}
with $x_i$ and $\sigma_i$ denoting the coordinates and internal quantum
numbers of the quasiparticles, and $N_{qp}$ their number. The above
Hamiltonian acts on many-particle wave functions
$\psi_{\{\sigma_i\}}(\{x_i\})$, which are bosonic (invariant under exchanges
$(x_i,\sigma_i)\leftrightarrow (x_j,\sigma_j)$), and correspond to states of
the form $|\psi) =\sum_{\{x_i\}}\sum_{\{\sigma_i\}}
\psi_{\{\sigma_i\}}(\{x_i\}) |\{x_i\},\{\sigma_i\})$.  
The dots in eq.~\eqref{eq:Heff} denote higher order terms, which
are irrelevant in the renormalization group sense, and do not influence the
asymptotic low-energy properties of the theory.

The scattering of two quasiparticles on each other can be
characterized by the two-particle S-matrix, which, in view of the
energy and momentum conservation, has a simple structure. The two-particle S-matrix, in
particular, relates the amplitude of an incoming asymptotic wave
function 
$\psi_{k_1\sigma_1,k_2\sigma_2}(x_1\ll x_2) \approx  A_{\sigma_1,\sigma_2}^{\rm in}(k_1,k_2) e^{i (k_1x_1 +k_2x_2)}$
with quasiparticle momenta $k_1>k_2$ to that of the outgoing wave function,
$\psi_{k_1\sigma_1,k_2\sigma_2}(x_1\gg x_2) \approx B_{\sigma_1,\sigma_2}^{\rm out}(k_1,k_2) e^{i (k_1x_1 +k_2x_2)}$ as
\begin{equation} 
{\bf B}^{\rm out} = {\hat S} (k_1-k_2) \;{\bf A}^{\rm in}\;.
%G
\end{equation}
The structure of the two-body S-matrix is further restricted 
by $\Z_3$ symmetry:
\begin{equation}
{\hat S} (k) = 
\left(\begin{array}{cccc}
s_3(k)& 0 & 0 & 0 \\
0& s_1(k) & s_2(k) & 0\\
0& s_2(k) & s_1(k) & 0\\
0& 0 & 0 & s_3(k)
\end{array}\right).
\end{equation}
In the following, we shall only investigate the scattering of quasiparticles in
the channels $\{+-\}$ and $\{-+\}$. In these channels, the eigenvalues of the
S-matrix read
\begin{eqnarray}
	s_t(k) &\equiv& e^{2 i \delta_t(k)} = s_1(k) + s_2(k)\;, \\
	s_s(k) &\equiv& e^{2 i \delta_s(k)} = s_1(k) - s_2(k)\;,
\end{eqnarray}
where we introduced the "triplet" and "singlet" eigenvalues, $s_t(k)$ and
$s_s(k)$, and the corresponding phase shifts, $\delta_t(k)$ and
$\delta_s(k)$. As shown in Ref.~\cite{qPotts}, interactions in the singlet
channel are irrelevant for $k\to0$ (the wave function has a node at
$x_1=x_2$), while they are relevant in the triplet channel, unless some very
special conditions are met by the effective interactions~\cite{qPotts}. As a
result, generically one finds $s_t(k\to0)=-1$ while $s_s(k\to0)=1$, as also
confirmed by direct calculations in the $g\to\infty$ and $g\to0$
limits~\cite{qPotts}. As a consequence, by analyticity, the phase shifts must
have the following small momentum expansion:
\begin{equation}
   \delta_t(k) = - \frac{\pi}{2}\;{\rm sgn}(k)  + a_t k +\dots\;, 
\phantom{nnn}\delta_s(k) = - a_s k + \dots\;. \label{eq:phase_shifts}
\end{equation}
Notice that these expressions (together with $s_3(k\to0)=-1$) give rise to a
low-momentum scattering matrix of the form, $\hat S \approx - \hat
X$. In contrast, perturbed conformal field theory yields a diagonal
low-momentum S-matrix with $s_t(k\to0)= 1$, corresponding to irrelevant 
interactions even in the  triplet channel. However, this would require very special 
interactions, and is not guaranteed by $\Z_3$ symmetry.

\subsection{Two-particle spectra: paramagnetic phase}
\label{su:para}

The two-particle spectrum of a finite system of size $L\gg a\equiv 1$
follows from the asymptotic form of the S-matrix. In the following, we
shall focus exclusively on the simplest case of periodic boundary
conditions (PBC). 

In the paramagnetic phase, quasiparticles carry a
``chirality'' label, $\sigma= Q=\pm$. Therefore, the $Q=+$ sector of
the spectrum contains single quasiparticle excitations 
of chirality $\sigma=+$ [described by
  eq.~\eqref{eq:e(k)}] as well as, e.g., two-particle excitations with
charges $\sigma_1=\sigma_2=-$. 
As a consequence, in the $Q=\pm$ sectors it is numerically hard to
separate two-particle states from the single-particle states. 
We therefore focus on the sector $Q=0$, where single quasiparticle
states are absent, and, above the ground state, the spectrum starts
directly with two-particle eigenstates of quasiparticles with charges
$\sigma_1=\pm$ and $\sigma_2=\mp$. 

\begin{figure}
	\centering
        \includegraphics[width=0.4\textwidth,clip=true]{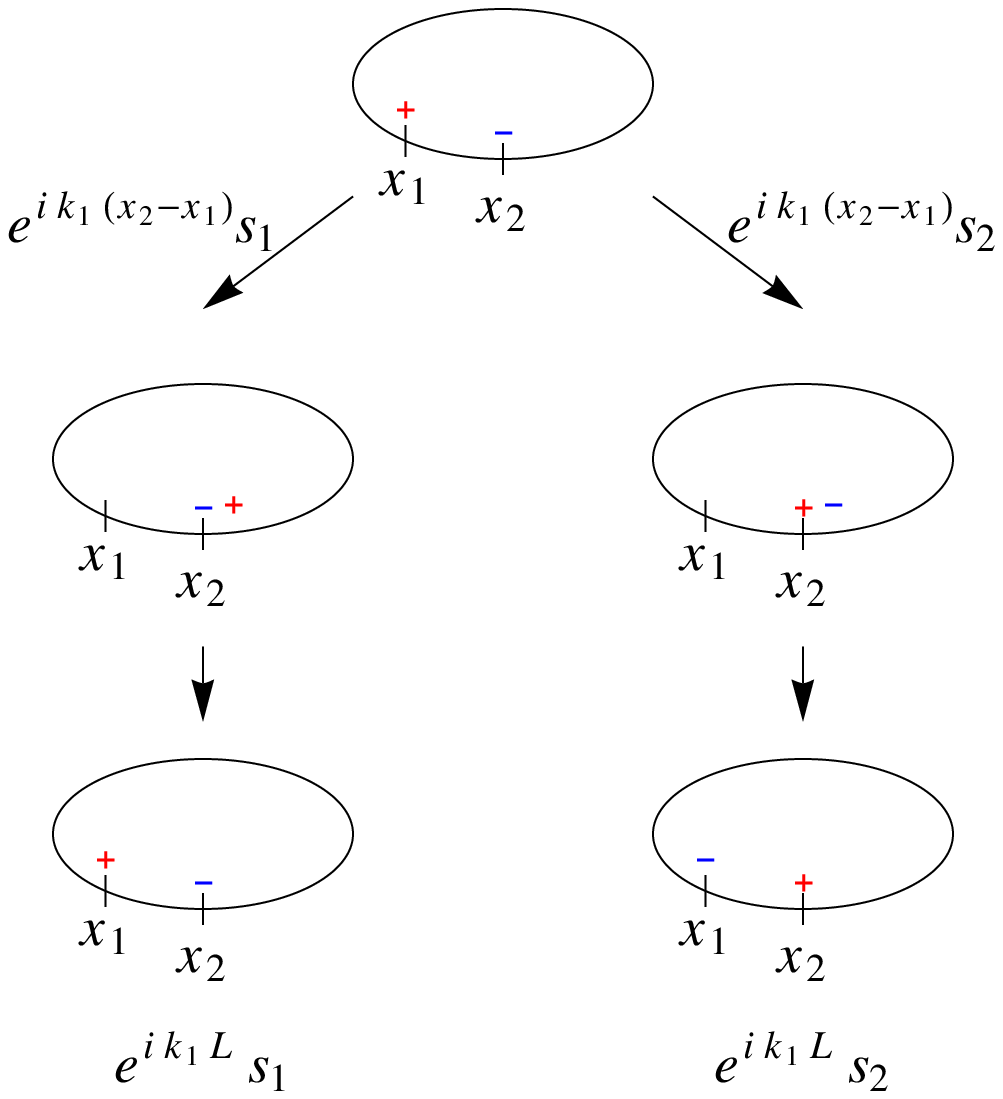}\hspace{2mm}
        \includegraphics[width=0.4\textwidth,clip=true]{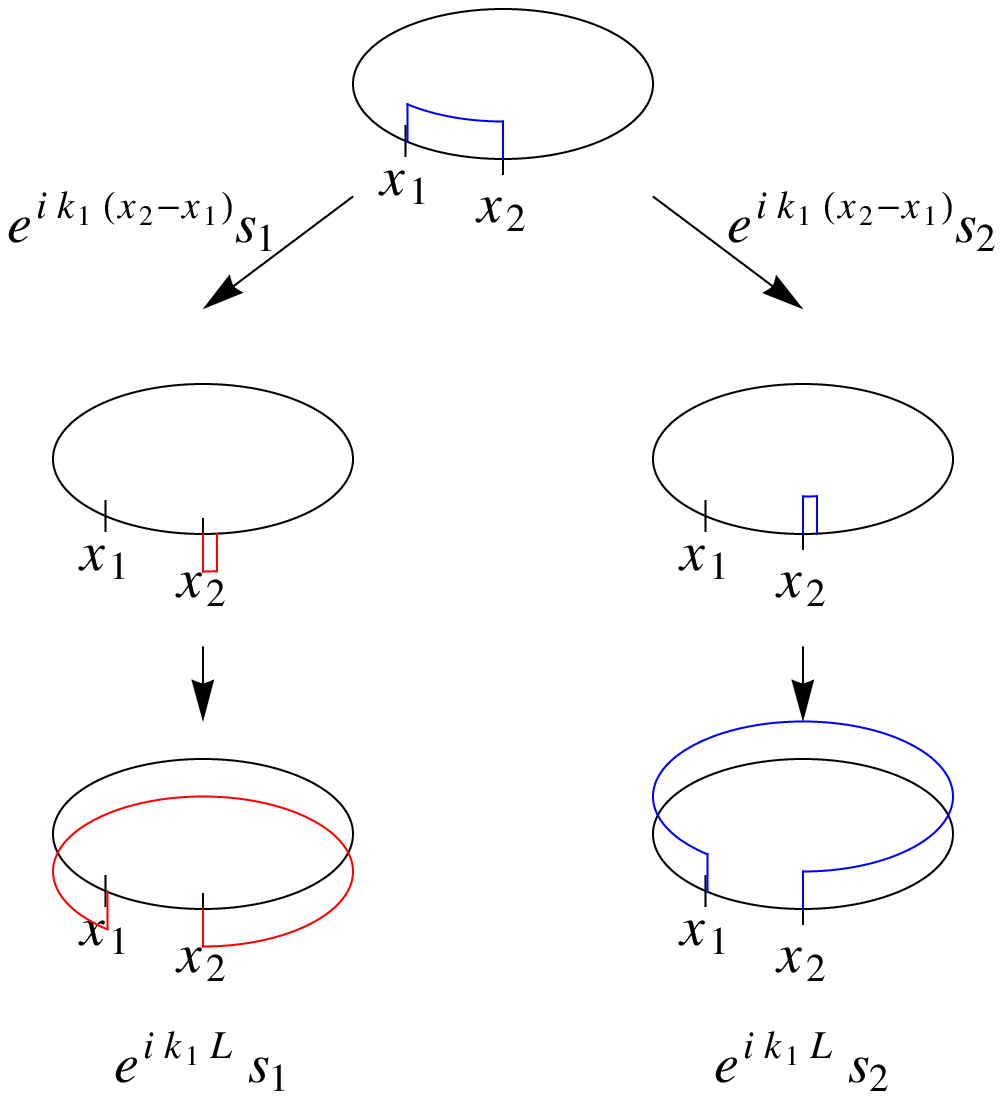}
	\caption{ \label{fig:roundP} \CO  Translating one of the quasiparticles around by 
system size $L$ in the paramagnetic (left figure) and 
ferromagnetic (right figure) phases. }
\end{figure}

For large system sizes, 
the quantization of the momenta $k_1$ and $k_2$ of the quasiparticles
is determined by the periodicity condition on
 the wave function, $\Psi(x_1,x_2)
\equiv \Psi(x_1+L,x_2) \equiv \Psi(x_1,x_2+L)$, and, just as in Bethe
Ansatz,   the energy is the sum of the two quasiparticle energies,
$E=\epsilon(k_1)+\epsilon(k_2)$. Taking particle $i=1$ around the
system (see fig.~\ref{fig:roundP}) then yields the following
condition, 
\begin{equation}
	\left(\begin{array}{c} C_1 \\ C_2  \end{array}\right) =
        e^{i k_1 L} \left(\begin{array}{cc} s_1(k_1-k_2) & s_2(k_1-k_2) \\ 
	s_2(k_1-k_2) & s_1(k_1-k_2)  \end{array}\right) 
	\left(\begin{array}{c} C_1 \\ C_2  \end{array}\right)  \;,
%G
\label{eqn:paramagn_quantcond}
\end{equation}
with $C_1=A_{+-}^{\rm in}(k_1\ge k_2)$ and $C_2=A_{-+}^{\rm
  in}(k_1\ge k_2)$ the wave function amplitudes for $0<x_1<x_2<L$, 
defined earlier. Taking particle $i=2$ around, $x_2\to x_2+L$, yields a
similar equation. 
In the triplet channel, $C_1=C_2$, we thus obtain
\begin{equation}
%G
	s_t(k_1-k_2) = e^{- ik_1L}  \;, \phantom{nnn}
s_t(k_2-k_1) = e^{- ik_2L} \;.
\label{eq:BA_triplet}
\end{equation}
Using the asymptotic expansions of the phase
shifts, 
eq.~\eqref{eq:phase_shifts}, and solving eq.~\eqref{eq:BA_triplet}
to leading order in $1/L$ then gives
\begin{equation}
E_{n_1,n_2}^t - 2\,\Delta = \Eo \; \left[\frac{1}{4} (n_1 + n_2)^2 +
  \frac{1}{4} \frac{(n_1-n_2+1)^2}{(1 + \frac{4
      a_t}{L})^2} + {\cal O}(1/L^2)\right]\;,
\label{eq:Etrip} 
\end{equation}
where $n_1$ and $n_2$ denote integers, and  
we introduced the energy unit,
\begin{equation}
\Eo \equiv \frac{ 1} {m}\left(\frac{2\pi}{L}\right)^2 \;.
\end{equation}
In eq.~\eqref{eq:Etrip}, to comply with the bosonic nature of the
excitations, the quantum numbers $n_1$ and $n_2$ must 
satisfy $n_1\ge n_2$. 

The previous analysis can be carried over to the
singlet sector, $C_1=-C_2$, with little modification, and  there
it yields the following finite size spectrum: 
\begin{equation}
E_{n_1, n_2}^s- 2\,\Delta =  \Eo \;\left[\frac{1}{4} (n_1 + n_2)^2 +
  \frac{1}{4} \frac{(n_1-n_2)^2}{(1 - \frac{4
      a_s}{L})^2} + {\cal O}(1/L^2)\right].\label{eq:Esing} 
\end{equation}
However, now $n_1$ and $n_2$ must satisfy  $n_1 > n_2$ since 
for  $n_1=n_2$ the wave function vanishes trivially.

\subsection{Two-particle spectra: ferromagnetic phase}

As discussed earlier, 
the ground state of the infinite system in the ferromagnetic phase has
broken $\Z_3$ symmetry, and correspondingly, it is 3-fold degenerate,  
$|0)_\mu$. Here we use brackets rather than angular brackets, to explicitly
emphasize that the states $\vert0 )_\mu$ are 
interacting many-body eigenstates of the Hamiltonian.
% \footnote{Here we used brackets rather than angular brackets, to explicitly
% emphasize that the states $\vert0 )_\mu$ are 
% interacting many-body eigenstates of the Hamiltonian.} 
Excitations are kinks (domain walls), and the corresponding 
two-particle states read
\begin{equation}
|x_1\,\theta_1, x_2\,\theta_2)_\mu\;,
\end{equation}
with $\mu$ the vacuum polarization at $x \to - \infty$, $x_i$ 
the positions of the domain steps, and $\theta_i=\pm$  the step
sizes. 
As we shall also demonstrate later through our 
finite size spectrum analysis, by duality, the S-matrix of 
these kinks is identical  to that of the local spin flip 
excitations on the paramagnetic side at a corresponding coupling, 
$g\to 1/g>1$. 

 On a ring,  PBC implies that $\theta_1+\theta_2=0$.  
Furthermore, in contrast to  the paramagnetic phase, in the ferromagnetic phase 
one must take  into account the presence of the 
three possible vacuum states when constructing periodic solutions. 
A way to do that is by keeping track of the vacuum polarization at 
position $x=0$, e.g.  As a consequence, 
 wave function amplitudes  must also have a vacuum label on the ring, 
$A_{\theta_1\theta_2}\to A_{\theta_1\theta_2}^{(\mu)}$. 
However, there is a subtle difference between
scattering in an infinite system and scattering on the ring. As
illustrated in fig.~\ref{fig:roundP}, moving one of the kinks around
results not only in a phase change and a collision of the elementary
excitations, but the domain orientations also change in a peculiar manner:
the configuration essentially turns ``inside out''. Correspondingly,
the amplitudes of the wave functions change as 
\begin{equation}
A_{+-}^{(\mu)}\to  e^{ik_1 L}\,s_1(k_1-k_2)\; A_{+-}^{(\mu-1)} + e^{ik_1 L}\,s_2(k_1-k_2)\;A_{-+}^{(\mu+1)}\;.
\label{eq:scattering_amps:FM}
\end{equation}

The states discussed so far are not eigenstates of the cyclic operator, $\mathcal{Z}$ (cf. eq.~\eqref{eq:C}). However, 
we can define eigenstates of $\mathcal{Z}$  by taking linear combinations of them. 
Combining e.g. the  three ferromagnetic ground states we find 
\begin{equation}
   \vert Q ) = \frac{1}{\sqrt{3}} \sum_\mu e^{-i \Omega Q \mu} 
\vert 0)_\mu\;. 
\label{eq:mixQ}
\end{equation}
Similarly, we can define the two-particle states,
 $|x_1\,\theta_1, x_2\,\theta_2;Q)$, and the corresponding
scattering states and wave function amplitudes, $A^{Q}_{\theta_1\theta_2}$, 
by simply mixing the states and the  wave function amplitudes 
as in  eq.~\eqref{eq:mixQ}.
Since the quantum number $Q$ is conserved, the periodicity condition 
of two-particle states  simplifies in this basis.  Taking the kink
$i=1$ around the ring, the relation in
eq.~\eqref{eq:scattering_amps:FM} implies the following 
equation for the amplitudes $C_1\equiv A_{+-}^Q$ 
and $C_2 \equiv A_{-+}^Q$ ,
\begin{equation}
  \left(\begin{array}{c} C_1 \\ C_2  \end{array}\right) = 
        e^{i k_1 L} \left(\begin{array}{cc} s_1(k_1-k_2) e^{-i\Omega Q} & s_2(k_1-k_2)
          e^{-i\Omega Q} \\ s_2(k_1-k_2) e^{i\Omega Q} & s_1(k_1-k_2) e^{i\Omega
            Q} \end{array}\right)    
\left(\begin{array}{c} C_1  \\ C_2  \end{array}\right)\;, \label{eq:phaseshifts_ferro} 
\end{equation}
and a similar equation is obtained for moving around particle
$i=2$. The structure of these equations is analogous to those in the
paramagnetic case, but the scattering lengths are replaced by
some effective $Q$-dependent scattering lengths, $a_{t,s}\to b^Q_{t,s}$,
yielding the triplet and singlet spectra 
 \begin{eqnarray}
	E_{n_1,n_2}^{Q,t}- 2\,\Delta &=&   \Eo \;\left[ \frac{1}{4} (n_1 +
          n_2)^2 
+ \frac{1}{4} \frac{(n_1-n_2+1)^2}{(1 +  \frac{4 b_t^{Q}}{L})^2} 
+ {\cal O}(1/L^2)
 \right]\;, \nonumber \\
	E_{n_1,n_2}^{Q,s}- 2\,\Delta  &=& \Eo \;\left[ \frac{1}{4}
          (n_1 + n_2)^2 + \frac{1}{4} \frac{(n_1-n_2)^2}{(1 - \frac{4
              b_s^Q}{L})^2} 
+ {\cal O}(1/L^2)
\right]. 
\label{eq:ferroenergies} 
\end{eqnarray}
Here the lengths $b_s^Q$ and $b_t^Q$ can be obtained by expanding the phases of
the eigenvalues 
of the matrix in eq.~(\ref{eq:phaseshifts_ferro}), 
\begin{equation}
	s_{s,t}^{Q}(k) \equiv e^{2 i \delta_{s,t}^Q(k)} = s_1(k) \cos
        \Omega Q \mp 
 \sqrt{s_2^2(k)- s_1^2(k) \sin^2 \Omega Q} \;,
\end{equation}
for low momenta. In the $Q=0$ sector the scattering lengths are thus given by $b_t^{Q=0} = a_t$  and
$b_s^{Q=0} = a_s$. The finite size spectrum  is thus 
in agreement with the duality relation,  eq.~\eqref{eq:duality}, 
provided that 
\be
a_t(g) = a_t(1/g)\;,\phantom{nnn}a_s(g) = a_s(1/g)\;.
\label{eq:duality_scatteringlength}
\ee
In the $Q=\pm1$ sector we get, on the other hand, 
\begin{equation}
	b_t^{Q=\pm 1} = \frac{1}{4} a_t - \frac{3}{4} a_s \;, b_s^{Q=\pm 1} = \frac{1}{4} a_s - \frac{3}{4} a_t \;. \label{eq:aQ1}
\end{equation}
Equations (\ref{eq:ferroenergies}) and (\ref{eq:aQ1}) are the most
important predictions of the effective theory. Together with the
duality relation between the ferromagnetic and paramagnetic phases,
they allow us to fit the DMRG data on the ferromagnetic side $g<1$
without any free fitting parameter. 

\section{Numerical results} 

\subsection{Technical details}

% Mode    N_Cut  NumSweeps   NumStates  
% 1       2000    5           3     
% 2       2500    5           7      
% 3       3000    5           11      
% 4       4000    5           15      
% 5       4500    5           15     
% 6       5000    5           15     

In the numerical calculations, we find the lowest-lying 
eigenstates of the Hamiltonian eq.~\eqref{eq:def:H} using 
the lattice units $J=a=\hbar=1$. 
We perform a standard DMRG calculation~\cite {DMRG}, where 
we make use of the
%G Abelian 
$\Z_3$ symmetry in the Q sector 
to perform a sub-blocking of the vector 
space.
More precisely, only the third order cyclic subgroup generated by
$\mathcal{Z}$ is exploited in the actual DMRG calculations.
% \footnote{
% More precisely, only the third order cyclic subgroup generated by
% $\mathcal{Z}$
% is exploited in the actual DMRG calculations.}
We use a two-site $A \bullet \bullet B$  super block configuration and 
targeted for up to 15 states lowest in energy. 
In order to achieve convergence for the larger system sizes,
we start with an initial run of targeting the lowest 3 states only,
performing 11 finite lattice sweeps and keeping 2000 states per $A$/$B$ block. 
We then restart this run increasing the number of low lying target
states to 7 and continue with 11, 15 low lying states keeping  
2500, 3000, and 4000 states per $A$/$B$ block performing 5 finite lattice sweeps in each restart.
For the 240 %and 500 
site systems we continued restarting the DMRG runs
keeping the 15 states lowest in energy and up to $5000 \ldots 7000$ 
states per block. In order to deal with the degeneracies and the large
number of low lying states we use the generalized Davidson algorithm 
ensuring that the solution of the preconditioner $u_n$ is orthogonal
to the previously found state $u_{n-1}$ which helps to avoid 
stagnation of the Davidson algorithm without paying the full overhead
of a Jacobi-Davidson scheme.   Calculations were performed with a
multi threaded code running on eight core machines with 64GB of RAM.  

\subsection{The central charge}

We can obtain the central charge of the critical theory by fitting the entanglement entropy  of a subsystem of size $x$,  $ S = - \mathrm{Tr} \rho_x \log(\rho_x) $ for $g=1$ by the result of Cardy and Calabrese \cite{entropy}
\[
    S_L(x) = \frac{c}{3}
    \log\left( \frac{\sin( \pi x / L ) L}{\pi} \right) + A
    \, .
\]
Here   $L$ is the total number of sites, $\rho_x$ stands for the reduced
density matrix of the subsystem, and $A$ is a non-universal offset. In this
way, we obtained for $L=120$ a conformal anomaly of $c=0.80042$. By performing
an $1/L$ fit to the  results obtained for system sizes between $L=12$ and
$L=120$, we obtained $c=0.799925$, in excellent agreement with the exact
result, $c=4/5$. 

\subsection{Single-particle levels}
\begin{figure} [t]
	\centering
	\includegraphics[width=0.65\textwidth,clip=true]{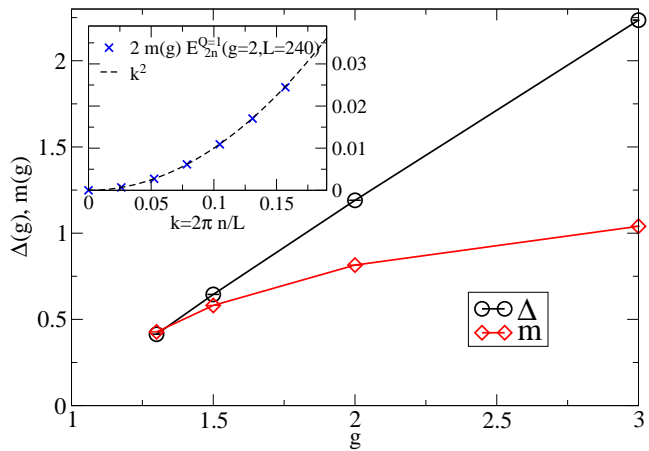}
	\caption{\CO  Single-particle parameters extracted from the
          DMRG data. Circles denote the quasiparticle gap
          $\Delta(g)$, while diamonds correspond to the
          quasiparticle mass $m(g)$. Units of $J=a=\hbar=1$ are used. 
           The standard error estimated for
          the fitting is less than the linewidth. Inset: quasiparticle
          dispersion relation $\epsilon_n - \Delta$ as a function of
          the momentum $k_n=2\pi n/L$. } 
	\label{fig:spparams}
\end{figure}

The numerically implemented  PBC forbids single quasiparticle excitations on the ferromagnetic 
side, where they can appear only under twisted boundary conditions. We note that domain wall excitations of 
the ferromagnetic phase  have  a $\Z_3$  charge $Q =0$.
% \footnote{Domain wall excitations of 
% the ferromagnetic phase  have  a $\Z_3$  charge $Q =0$.} 
In contrast, in the paramagnetic phase the charge of 
quasiparticles is $Q=\pm$.  Therefore, while we could not investigate 
single quasiparticle excitations in the ferromagnetic phase, 
we could study them in the paramagnetic phase in the $Q=\pm$ sectors, where
they  appear as the lowest-lying excitations. 
Equation (\ref{eq:e(k)}) and PBC imply in the $Q=\pm$ sectors of the paramagnetic phase
that, for very large systems, the single-particle energies are given by 
\begin{equation}
\epsilon_n (g>1) = \Delta(g) + \frac{1}{2 m(g)} \left( \frac{2\pi}{L} n
\right)^2  + \dots, \label{eq:spectr_p1}
\end{equation}
with $n \in {\mathbb Z}$.  The quasiparticle gap can thus be 
identified  as
\begin{equation}
\Delta(g > 1) \equiv \lim_{L \to \infty} \left( E^{Q=1}_{n=0}(g,L) -
E^{Q=0}_{n=0}(g,L)
 \right)\;,
\end{equation}
and can be obtained from extrapolating the corresponding numerical
data to  $L^{-1} \to 0$. The quasiparticle mass $m(g)$ 
can be defined and extracted through a similar extrapolation procedure.

The single-particle parameters obtained this way are shown in
fig.~\ref{fig:spparams}. The inset demonstrates that the quadratic
dispersion is indeed consistent with the numerically computed excitation spectrum.
Both $m$ and $\Delta$ decrease as the coupling
approaches the critical value, $g \to 1$, where the gap is supposed to
vanish as $\Delta \sim J |g-1|^{5/6}$. The data 
are consistent with this power-law behavior, but it is difficult to
extract the precise value of the critical exponent from them. 

The fact that due to the lack of the single-particle excitations, we
cannot  obtain the quasiparticle parameters
on the ferromagnetic side, $g<1$, directly from the DMRG data is of
little concern. 
The duality relation, eq.~\eqref{eq:duality} relates the 
quasiparticle gaps and masses in the
 two phases, since for two remote quasiparticles in a very large system 
we must have
%for the two-particle levels
\begin{equation}
\epsilon(k_1,g)+\epsilon(k_2,g) =  g \; (\epsilon(k_1,1/g)+\epsilon(k_2,1/g))\;.
\end{equation}
This can hold for all momenta $k_1,k_2$ only if
\begin{equation}
 m(g) = \frac 1 g \; m(1/g)\;,\phantom{nn}  \Delta(g) = g \; \Delta(1/g)\;.
\label{eq:massduality}
\end{equation}

\subsection{Two-particle levels: paramagnetic phase}

\begin{table}
\caption{
\label{tbl:asymptotics}
a) Asymptotic values of normalized two-particle
  energies, $\epsilon_{n_1,n_2}\equiv (E_{n_1,n_2}-2\Delta)/\Eo$,
for $L \to \infty$ as predicted by the effective Hamiltonian for a
reflective S-matrix in the $Q=0$ sector of the
paramagnetic phase, and the corresponding rescaled energy values, from DMRG (without extrapolation to $L\to \infty$). The DMRG data are
taken at $g=2$ for $L=240$. b) asymptotic values of energy levels assuming a diagonal S-matrix.}

\begin{indented}

\vspace{5mm}

\item[a)] \hspace{3mm} \begin{tabular}{@{}llll}
\br
$\epsilon^{}_{n_1,n_2}$ & $(n_1,n_2)$ & parity &
DMRG \\
\mr
1/4 & (0,0) & t &  0.21995
(1$\times$) \\

1/2 & (1,0),(0,-1) & s &
0.50658 (2$\times$) \\

1 & (1,-1) & s &
1.02647 (1$\times$) \\

5/4 & (1,0),(0,-1) & t
& 1.13124 (2$\times$) \\

5/4 & (1,1),(-1,-1)
& t &  1.21941 (2$\times$) \\
\br
\end{tabular}

\vspace{5mm}

\item[b)] \hspace{3mm}  \begin{tabular}{@{}lll}
\br
$\epsilon^{\rm diag}_{n_1,n_2}$ & $(n_1,n_2)$ &
parity \\
\mr
0 & (0,0) & t  \\

1/2 & (1,0),(0,-1) & s \\

1/2 &
(1,0),(0,-1) & t \\

1 & (1,-1) & s \\

1 & (1,-1) & t \\

1 &
(1,1),(-1,-1) & t \\
\br
\end{tabular}
\end{indented}
\end{table}

The prediction of our effective field theory is that for very 
large system sizes, $L\to\infty$, 
the excitation spectrum becomes universal 
in the sense that the rescaled energies, 
$\epsilon_{n_1,n_2}\equiv (E_{n_1,n_2}-2\Delta)/\Eo$, 
approach  universal fractions and have corresponding 
universal degeneracies. Furthermore, eqns.~\eqref{eq:Etrip},\eqref{eq:Esing} 
 and   \eqref{eq:ferroenergies} also predict that
corrections to this universal spectrum  can be fitted just in 
terms of two scattering lengths,  $a_s$ and $a_t$ for all levels. 
The predicted universal spectrum, and its comparison with the 
numerically obtained finite size
spectrum is shown in Table~\ref{tbl:asymptotics}.a. Already without
incorporating finite size corrections, a very good agreement is 
found: all degeneracies as well as the approximate energies 
of the states agree very well with the predictions of
eqns.~\eqref{eq:Etrip} and \eqref{eq:Esing}. 
However, as we demonstrate in Table~\ref{tbl:asymptotics}.b, 
the finite size spectrum  is completely inconsistent with the spectrum 
associated with a diagonal S-matrix: there the phase shifts
$\delta^{s}_{\rm diag}$ and $\delta^{t}_{\rm diag}$ vanish for small momenta, 
and the asymptotic values of the normalized levels are given by
$(n_1+n_2)^2/4 + (n_1-n_2)^2/4$, with $n_1 \geq n_2$ and $n_1 > n_2$
for the triplet and singlet sectors, respectively. We remark
that the same (inconsistent) %(incorrect) 
values are given by the perturbed CFT calculations, discussed in Section~\ref{sec:scalingPotts}. 

\begin{figure}
	\centering
	\includegraphics[width=0.65\textwidth,clip=true]{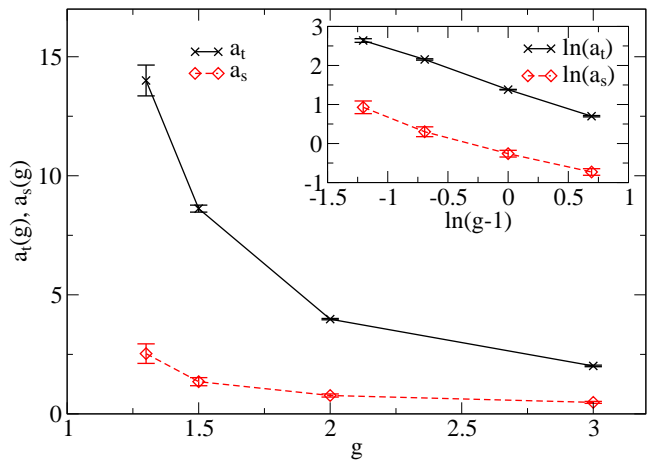}
	\caption{\CO
\label{fig:atas}
 Scattering lengths $a_t(g)$ and $a_s(g)$ as a function of $g$ in
 units of the lattice constant, $a$. Inset: 
the logarithms of the scattering lengths as functions of $\ln (g-1)$ show a power law divergence. }
\end{figure}

An even more consistent picture based on the effective theory is obtained if one also incorporates 
$1/L$ corrections due to the finite scattering lengths, 
$a_s$ and $a_t$. The latter quantities can be extracted from the 
finite size spectrum by using only the lowest two excited states 
in the $Q=0$ subspace 
%the scattering lengths are defined as 
\bea
	a_t(g>1) &= &- \lim_{L\to\infty} \frac{L}{2} \left[
          \left(E_{n=1}^{Q=0}(g,L) - 2\Delta(g) \right)/\Eo(g,L) -
          \frac{1}{4}\right]
\;,
\\
	a_s(g>1) &=& \phantom{-} \lim_{L\to\infty}  \frac{L}{2} 
\left[ \left(E_{n=2}^{Q=0}(g,L) - 2\Delta(g) \right)/\Eo(g,L) - \frac{1}{2} \right],
\eea 
respectively. The $g$-dependence of the 
extracted scattering lengths, $a_s(g)$ and $a_t(g)$ is 
shown in fig.~\ref{fig:atas}.  Similar to the correlation length, 
$a_s$ and  $a_t$ both seem to diverge at the
critical point, but it is not possible to extract an
accurate exponent from our numerical data.  

\begin{figure}
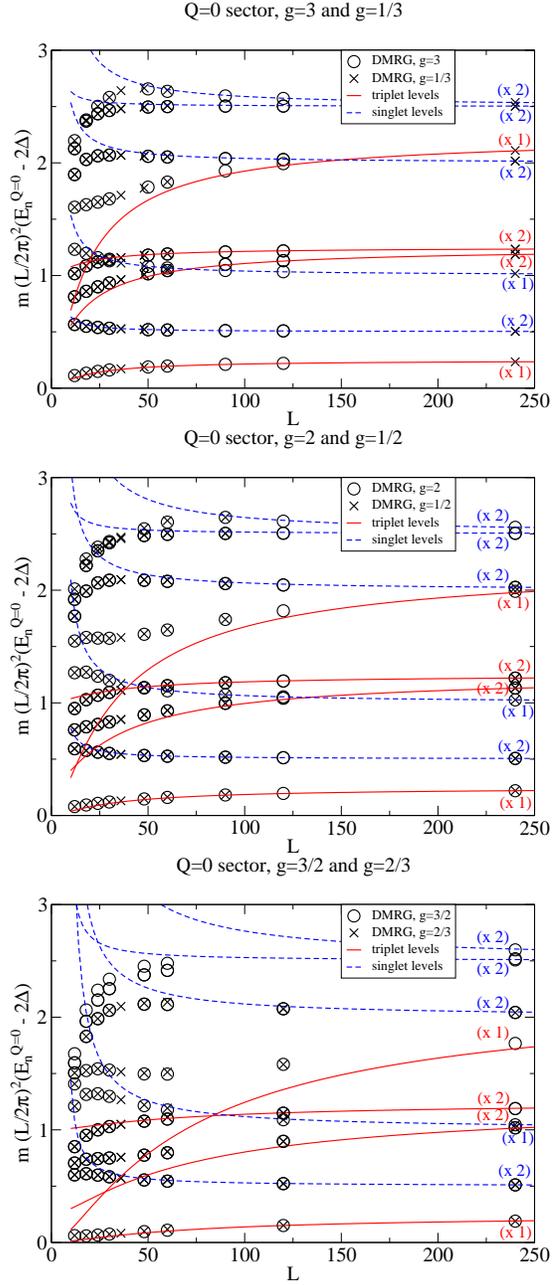

	\centering
\includegraphics[width=0.55\textwidth,clip=true]{dual-g3.eps}
\includegraphics[width=0.55\textwidth,clip=true]{dual-g2.eps}
\includegraphics[width=0.55\textwidth,clip=true]{dual-g1.5_v2.eps}
	\caption{\CO
\label{fig:Q0spectra} 
Comparison of the rescaled energies from DMRG and the effective theory
for $g = 3$, $g = 2$, and $g = 3/2$ in the $Q=0$ sector. The scattering lengths $a_t$ and
$a_s$ were fitted using the lowest lying two levels, respectively. No
further fitting for the higher levels was used. Using the
duality of the model, we also show the rescaled ferromagnetic spectrum for $\tilde
g=1/g$ in the $Q=0$ sector.}
\end{figure}

Although the extrapolations in the case of the scattering lengths are
less accurate than in the case of the single-particle parameters, we
can now plot the higher energy levels in the $Q=0$ sector using
eqns.~(\ref{eq:Etrip}) and (\ref{eq:Esing}), and compare them to the
appropriately rescaled DMRG data for various system sizes. 
We find a very good agreement between the numerical results 
and the predictions for the spectrum of the effective model, 
as can be seen in fig.~\ref{fig:Q0spectra}. A clear convergence to the
asymptotic values is observed for large $L$'s, 
and deviations appear only at smaller system sizes
or at  higher energy levels, where the asymptotic description must 
break down. 

\subsection{Two-particle levels: ferromagnetic  phase}

As we discussed before, the spectrum on the ferromagnetic side $g<1$
with PBC does not have any single-particle levels, from which we could get the
quasiparticle parameters directly. Although, in
principle, it would be possible to fit the quasiparticle gap and
quasiparticle mass along with the scattering lengths from the two-body
spectra given by eqns.~\eqref{eq:ferroenergies}, this is not needed. 
As discussed earlier, the duality relation, eq.~\eqref{eq:duality}
connects energy scales for $g\leftrightarrow 1/g$, 
and thus $\Delta(g)$ and $m(g)$ through eq.~\eqref{eq:massduality} 
in both phases. In addition, it also implies 
that all length scales emerging in the problem must  be
invariant under the duality transformation $g\leftrightarrow 1/g$:   
relevant length scales appear in the finite size spectrum as
cross-over scales, and  by the invariance of the spectrum, they must 
transform similar to the scattering lengths,  
eq.~\eqref{eq:duality_scatteringlength}.

Fig.~\ref{fig:Q0spectra} provides an explicit numerical evidence for
the duality relation in the $Q=0$ sector. 
There we show that the appropriately rescaled DMRG
data obtained for $g<1$ completely overlap with the data 
on  the paramagnetic side $g\to 1/g$. This gives 
a numerical proof for the relations \eqref{eq:duality}, \eqref{eq:duality_scatteringlength}, and \eqref{eq:massduality}.
We emphasize that the duality relation holds for all system sizes $L$. 
 
\begin{figure}
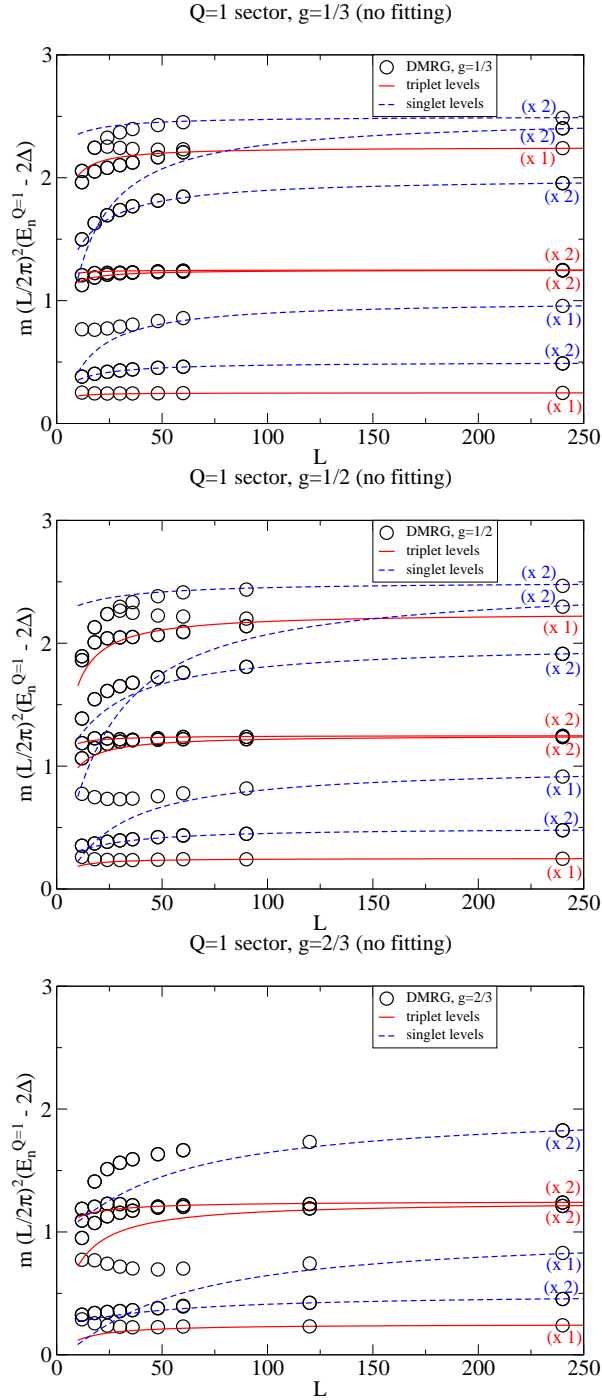

	\centering
	\includegraphics[width=0.6\textwidth,clip=true]{Q1-g3.eps}
	\includegraphics[width=0.6\textwidth,clip=true]{Q1-g2.eps}
	\includegraphics[width=0.6\textwidth,clip=true]{Q1-g1.5_v2.eps}
	\caption{\CO  
\label{fig:Q13}	
The spectra for $g=1/3$, $g=1/2$,$g=2/3$
 in the $Q=1$ sector. Only the numerically converged data points and the corresponding curves from the effective theory are shown. }
\end{figure}

In addition to the consistency between the effective
theory with a reflective asymptotic S-matrix and the numerical data in
the $Q=0$ sector, probably the most important check of validity of our
assumptions is the comparison to the DMRG results in the
ferromagnetic $Q=1$  sector. There the predicted  
finite size spectrum is  given by eqns.~\eqref{eq:ferroenergies}
together with eq.~\eqref{eq:aQ1}. We emphasize 
that all parameters of the effective theory are fixed already 
and no further adjustment to the
theoretical spectra is possible. As we show in fig.~\ref{fig:Q13},
the effective theory presented here also describes the numerical 
data in the $Q=1$ sector within numerical precision in the
regime, $\xi \ll L$, and fully confirms
  eqns.~\eqref{eq:ferroenergies} and \eqref{eq:aQ1}.
 Here $\xi = (m c)^{-1}$ is the correlation length, see also eq.~(\ref{eq:l}).

\section{Comparison to the scaling Potts field theory}
\label{sec:scalingPotts}

\subsection{Scaling Potts field theory as a perturbed conformal field theory \label{sub:Scaling-Potts-model}}

Here we only give a brief review of the scaling Potts field theory; a detailed
analysis of the scattering theory is given in a separate paper \cite{Takacs:2011xx}.
The scaling limit of the three-state Potts model at the critical point is a minimal
conformal field theory with central charge 
$C=\frac{4}{5}$ \cite{Belavin:1984vu,Dotsenk:1984}. The Kac table of
conformal weights is
\[
\left\{ h_{r,s}\right\} =\left(\begin{array}{ccccc}
0 & \frac{1}{8} & \frac{2}{3} & \frac{13}{8} & 3\\
\frac{2}{5} & \frac{1}{40} & \frac{1}{15} & \frac{21}{40} & \frac{7}{5}\\
\frac{7}{5} & \frac{21}{40} & \frac{1}{15} & \frac{1}{40} & \frac{2}{5}\\
3 & \frac{13}{8} & \frac{2}{3} & \frac{1}{8} & 0
\end{array}\right)\qquad{r=1,\dots,4;\atop s=1,\dots,5.}
\]
The sectors of the Hilbert space are products of the irreducible representations
of the left and right moving Virasoro algebras which can be specified
by giving their left and right conformal weights as 
\[
\mathcal{S}_{h,\bar{h}}=\mathcal{V}_{h}\otimes\mathcal{V}_{\bar{h}}\;.
\]
There are two possible conformal field theory partition functions for
this value of the central charge \cite{Cappelli:1986hf}. The one
describing the three-state Potts model is the $D_{4}$ modular invariant, for
which the complete Hilbert space is 
\begin{eqnarray}
\mathcal{H} & = & \mathcal{S}_{0,0}\oplus\mathcal{S}_{\frac{2}{5},\frac{2}{5}}\oplus\mathcal{S}_{\frac{7}{5},\frac{7}{5}}\oplus\mathcal{S}_{3,3}\nonumber \\
 &  & \oplus\mathcal{S}_{\frac{1}{15},\frac{1}{15}}^{+}\oplus\mathcal{S}_{\frac{1}{15},\frac{1}{15}}^{-}\oplus\mathcal{S}_{\frac{2}{3},\frac{2}{3}}^{+}\oplus\mathcal{S}_{\frac{2}{3},\frac{2}{3}}^{-}\nonumber \\
 &  &
\oplus\mathcal{S}_{\frac{2}{5},\frac{7}{5}}\oplus\mathcal{S}_{\frac{7}{5},\frac{2}{5}}\oplus\mathcal{S}_{0,3}\oplus\mathcal{S}_{3,0}
\;.
\label{eq:Dinvariantsectors}
\end{eqnarray}
Note that not all of the possible representations occur in the Hilbert
space; there is another modular invariant partition function called
$A_{4}$ which includes all sectors of diagonal form $\mathcal{S}_{h,h}$
allowed by the Kac table exactly once:
$\mathcal{H}=\bigoplus_{h}\mathcal{V}_{h}\otimes\mathcal{V}_{h}$.
The $A_{4}$ model corresponds to the scaling limit of a higher multicritical
Ising class fixed point with symmetry $\mathbb{Z}_{2}$. In contrast,
the $D_{4}$ conformal field theory is invariant under the permutation
group $\Z_{3}$ generated by two elements $\mathcal{Z}$ and
$\mathcal{C}$ with the relations
\[
\mathcal{Z}^{3}=1\;,\qquad\mathcal{C}^{2}=1\;,
\qquad\mathcal{CZC}=\mathcal{Z}^{-1}\;,
\]
which have the signatures
$\mbox{sign }\mathcal{Z}=+1$ and $\mbox{sign }\mathcal{C}=-1$.
The sectors in the first line of (\ref{eq:Dinvariantsectors}) are
invariant under the action of the permutation group, $\Z_{3}$,
while the two pairs on the second line each form 
 two-dimensional irreducible
representations, as characterized by the following action of
the generators:
\be
\mathcal{C}|\pm\rangle  =  |\mp\rangle\;,\quad
\mathcal{Z}|\pm\rangle  =  \mathrm{e}^{\pm i \Omega}|\pm\rangle\;.
\ee
Finally, sectors  in the third line of eq.~\eqref{eq:Dinvariantsectors}
form one-dimensional signature
representations, where each element is represented by its signature.
These sectors are in one-to-one correspondence with the families of
conformal fields: the primary field 
in the family corresponding to $\mathcal{S}_{h,\bar{h}}$
has left and right conformal weights $h$ and $\bar{h}$, and a
corresponding scaling dimension, $\Delta_{h,\bar{h}}= h +\bar{h}$,
and is denoted by $\Phi_{h,\bar{h}}$, with an optional upper $\pm$ index for
fields forming a doublet of ${\Z}_{3}$. 

Relevant fields are exactly those for which $h+\bar{h}<2$. It is
then obvious that the only $\mathbb{Z}_{3}$-invariant spinless relevant
field is $\Phi_{\frac{2}{5},\frac{2}{5}}$, 
which means that the Hamiltonian of the scaling limit of the off-critical
three-state Potts model is uniquely determined \cite{Dotsenk:1984},
\begin{equation}
H=H_{*}+\tau\int dx\;\Phi_{\frac{2}{5},\frac{2}{5}}
\;.
\label{eq:scaling_potts_hamiltonian}
\end{equation}
The sign of the coupling constant $\tau$ corresponds to the two phases:
$\tau>0$ is the paramagnetic, while $\tau<0$ is the ferromagnetic
phase. Up to normalization factor, it is given by
\begin{equation}
\tau\propto(g-1)a^{-6/5}\label{eq:tau_g}\;,
\end{equation}
with $a$ the lattice spacing. The scaling limit is achieved by
taking $a\rightarrow0$ and $g\rightarrow g_c=1$ such that $\tau$ remains finite. In this limit, the  gap 
\[
\Delta \sim \tau^{5/6}\sim |g-1|^{5/6} \;\hbar c/a
\]
remains also finite. Here, for clarity, we restored $\hbar$
and $c$, which are usually both set to unity in relativistic quantum field theory.

The scaling Potts field theory (\ref{eq:scaling_potts_hamiltonian}) is known
to be integrable \cite{Zamolodchikov:1987zf}, and its spectrum and
scattering matrix was determined
exactly~\cite{Zamolodchikov:1987zf,Pottsintegr}. 
In the paramagnetic phase, the vacuum is non-degenerate and the spectrum
consists of a pair of particles $A$ and $\bar{A}$ of mass $m$, 
 which
form a doublet under $\mathbb{Z}_{3}$ \cite{Smirnov:1991uw}: 
\begin{eqnarray}
\mathcal{C}|A(\beta)\rangle=|\bar{A}(\beta)\rangle\;, & \qquad &
\mathcal{Z}|A(\beta)\rangle=\mathrm{e}^{i \Omega}|A(\beta)\rangle\nonumber\;, \\ 
\mathcal{C}|\bar{A}(\beta)\rangle=|A(\beta)\rangle\;, & \qquad &
\mathcal{Z}|\bar{A}(\beta)\rangle=\mathrm{e}^{-i\Omega}|\bar{A}(\beta)\rangle\;.\label{eq:s3action-on-aabar} 
\end{eqnarray}
The excitations  $A$ and $\bar{A}$ correspond to the local spin flip
  excitations of chirality $\sigma=Q=\pm$ of the lattice.
The generator $\mathcal{C}$ is identical to charge conjugation ($\bar{A}$
is the antiparticle of $A$). Choosing units in which $\hbar=c=1$,
two-dimensional Lorentz invariance implies that the energy and momentum
of the particles can be parameterized by the rapidity $\beta$:
\[
E=m\cosh\beta,\quad p=m\sinh\beta\;.
\]
The two-particle scattering amplitudes are~\cite{Zamolodchikov:1987zf}
\begin{eqnarray}
S_{AA}(\beta_{12}) & = & S_{\bar{A}\bar{A}}(\beta_{12})=\frac{\sinh\left(\frac{\beta_{12}}{2}+\frac{\pi i}{3}\right)}{\sinh\left(\frac{\beta_{12}}{2}-\frac{\pi i}{3}\right)}\;,\nonumber \\
S_{A\bar{A}}(\beta_{12}) & = &
S_{\bar{A}A}(\beta_{12})=-\frac{\sinh\left(\frac{\beta_{12}}{2}+\frac{\pi
    i}{6}\right)}{\sinh\left(\frac{\beta_{12}}{2}-\frac{\pi
    i}{6}\right)}\;,
\label{eq:highT_Smatrix}
\end{eqnarray}
where $\beta_{12}=\beta_{1}-\beta_{2}$ is the rapidity difference
of the incoming particles. This $S$ matrix was confirmed by thermodynamic
Bethe Ansatz \cite{Zamolodchikov:1989cf}. We remark that the pole
in the $S_{AA}=S_{\bar{A}\bar{A}}$ amplitudes at
$\beta_{12}=\frac{2\pi i}{3}$
corresponds to the interpretation of the particle $\bar{A}$ as a bound
state of two particles $A$, and similarly, $A$ as a bound state of
two $\bar{A}$s, under the bootstrap principle (a.k.a. ``nuclear
democracy''). 
The pole in $S_{A\bar{A}}=S_{\bar{A}A}$ amplitudes at
$\beta_{12}=\frac{\pi i}{3}$ has a similar interpretation in the
crossed channel, and it does not correspond to a true bound state 
in the neutral sector.

The excitations in the ferromagnetic phase are topologically charged
\cite{Pottsintegr}. Similar to the lattice model, 
 the vacuum is three-fold degenerate $|0)_{\mu}$ ($\mu=-1,0,1$).
The action of  $\Z_{3}$ on the vacua is 
\[
\mathcal{Z}|0)_{\mu}=|0)_{\mu+1\bmod3}\;,\qquad\mathcal{C}|0)_{\mu}=|0)_{-\mu}\;,
\]
and the excitations are kinks of mass $m$ interpolating between adjacent
vacua, and correspond to domain walls on the lattice.
 The kink of rapidity $\beta$, interpolating from $\mu$ to
$\mu'$ is denoted by 
\[
K_{\mu\mu'}(\beta)\;,\qquad \mu-\mu'=\pm1\bmod3\;.
\]
The scattering processes of the kinks are of the form 
\[
K_{\mu \nu}(\beta_{1})+K_{\nu \mu'}(\beta_{2})\rightarrow K_{\mu \nu'}(\beta_{1})+K_{\nu' \mu'}(\beta_{2})\;,
\]
with the scattering amplitudes equal to
\begin{equation}
S\left(\mu{\nu'\atop \nu}\mu'\right)(\beta_{12})=
% \begin{cases}
\cases{ S_{AA}(\beta_{12}) \qquad {\rm if}  \qquad \nu =\nu' \;, \\
S_{A\bar{A}}(\beta_{12}) \qquad {\rm if} \qquad \mu =\mu'\;.
}
% \end{cases}
\label{eq:lowT_Smatrix}
\end{equation}
This essentially means that, apart from the restriction of kink succession
dictated by the vacuum indices (adjacency rules), the following identifications
can be made 
\be
K_{\mu\nu}(\beta)\equiv 
% \begin{cases}
 \cases{ A(\beta)\qquad \mu - \nu=+1\bmod3\;,\\
\bar{A}(\beta)\qquad \mu-\nu=-1\bmod3\;.
}
% \end{cases}
\ee
in all other relevant physical aspects (such as e.g. the bound state interpretation
given above). 

The validity of the S-matrix expressions
(\ref{eq:highT_Smatrix}) and (\ref{eq:lowT_Smatrix}) 
 for the scaling Potts model can be 
checked by comparing the finite size spectrum of the corresponding
Bethe Ansatz equations to that of \eqref{eq:scaling_potts_hamiltonian}
as obtained by the truncated conformal space approach (TCSA).
The TCSA was originally developed in \cite{Yurov:1989yu}.
% \footnote{
% The truncated conformal space approach was originally developed in
% \cite{Yurov:1989yu}.} 
This is performed in detail
in~\cite{Takacs:2011xx}. 
In the TCSA, one determines the finite size
spectrum of eq.~\eqref{eq:scaling_potts_hamiltonian}  numerically 
by truncating the finite volume
Hilbert space by imposing an upper cutoff in the eigenvalue of the
conformal Hamiltonian.  For the ground state, this is equivalent to
the standard variational 
calculus in quantum theory, where the variational wave function Ansatz
is expressed as a linear combination of a finite subset of the
eigenstates of the conformal Hamiltonian.
By looking at the conformal fusion rules implied by the three-point
couplings \cite{Fuchs:1989kz,Petkova:1988yf,Petkova:1994zs}, it turns
out that the perturbing operator acts separately in the following
four sectors:
\begin{eqnarray}
\mathcal{H}_{0} & = &
\mathcal{S}_{0,0}\oplus\mathcal{S}_{\frac{2}{5},\frac{2}{5}}\oplus\mathcal{S}_{\frac{7}{5},
  \frac{7}{5}}\oplus\mathcal{S}_{3,3}\;,\nonumber  
\\
\mathcal{H}_{\pm} & = &
\mathcal{S}_{\frac{1}{15},\frac{1}{15}}^{\pm}\oplus\mathcal{S}_{\frac{2}{3},\frac{2}{3}}^{\pm}
\;,
\nonumber \\
\mathcal{H}_{1} & = &
\mathcal{S}_{\frac{2}{5},\frac{7}{5}}\oplus\mathcal{S}_{\frac{7}{5},\frac{2}{5}}\oplus\mathcal{S}_{0,3}\oplus\mathcal{S}_{3,0}\;,\label{eq:Hilbert_space_sectors} 
\end{eqnarray}
so the Hamiltonian can be diagonalized separately in each of them.
In the lattice language,  $\mathcal{H}_{\pm}$ correspond to
  the  sectors $Q=\pm$, while $\mathcal{H}_{0} \oplus \mathcal{H}_{1} $
span the $Q=0$ sector. Charge conjugation $\mathcal{C}$  implies 
that the  Hamiltonian is exactly identical in the
sectors $\mathcal{H}_{+}$ and $\mathcal{H}_{-}$. Furthermore, 
the spectrum is invariant under  transformation, 
$\tau\rightarrow-\tau$ in sectors
$\mathcal{H}_{0}$ and $\mathcal{H}_{1}$. 
This  is the consequence of a $\mathbb{Z}_{2}$ symmetry in these
sectors, 
which  leaves the fixed point Hamiltonian $H_{*}$ and the
conformal fusion rules in these sectors invariant. We remark that the conformal fusion rules do not allow the extension of this symmetry to the $\mathcal{H}_{\pm}$. 
% \footnote{The conformal fusion rules do not allow the extension of this symmetry
% to the $\mathcal{H}_{\pm}$. }
Away from the critical point, it can be interpreted as the continuum
form of the duality transformation (\ref{eq:duality}) 
in the scaling limit. 

As further discussed in ~\cite{Takacs:2011xx}, 
the detailed TCSA calculations indeed confirm that the $S$-matrices
(\ref{eq:highT_Smatrix}) and (\ref{eq:lowT_Smatrix}) correctly describe
the paramagnetic and ferromagnetic phases of the scaling 
field theory. However, as we discuss below, the Bethe Ansatz
spectra computed with (\ref{eq:highT_Smatrix}) and
(\ref{eq:lowT_Smatrix}) both turn out to be  {\em inconsistent} 
with the numerically computed finite size spectra.

\subsection{Comparing the scaling field theory to DMRG}
\label{sec:DMRG_vs_scalingFieldTheory}

\begin{figure}
\begin{centering}
\includegraphics[scale=0.4,clip=true]{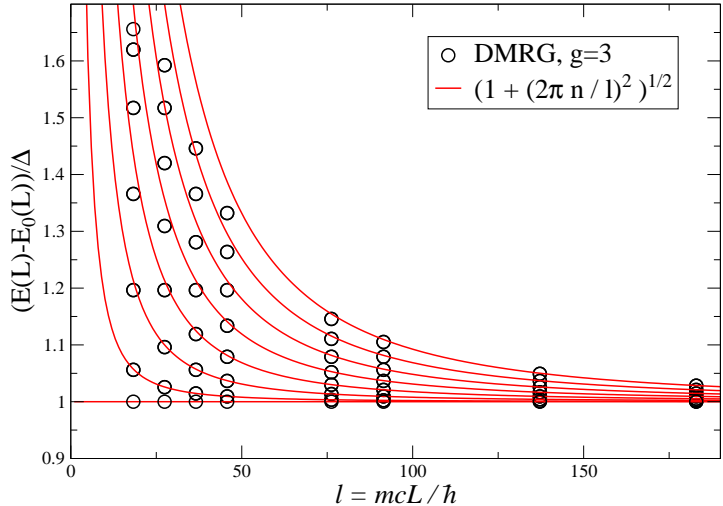}
\par\end{centering}
\caption{\label{fig:DMRGopt}\CO Comparing rescaled DMRG data (for $g=3$) to
the relativistic one-particle dispersion relation. Length is measured
in the Compton length, $l=mcL/\hbar$.}
\end{figure}

In order to compare the DMRG to the scattering matrices (\ref{eq:highT_Smatrix},\ref{eq:lowT_Smatrix})
directly, we need to rescale the variables to appropriate units in
which $c=1$. The relativistic relation 
\begin{equation*}
\Delta=mc^{2} 
\end{equation*}
allows to determine the speed of light $c = \sqrt{\Delta/m}$ 
in lattice units ($aJ/\hbar$). We recall that, 
according to eq. (\ref{eq:e(k)}), $\Delta$
is the infinite volume limit of the energy gap between the stationary 
one-particle state
and the ground state, while $m$ can be determined from the large
volume behavior of the first excited one-particle state. We then
introduce the dimensionless volume variable ($\hbar=1$)
\begin{equation}
l=mcL\;,  \label{eq:l}
\end{equation}
i.e. we measure the volume in units of the Compton length. 
After rescaling the DMRG spectrum to these units, we  expect 
the spectrum of one-particle states to follow the relativistic
dispersion, 
\begin{eqnarray*}
\frac{1}{\Delta}\left(E(L)- E_{0}(L)\right) & = &
\sqrt{1+\left(\frac{p}{mc}\right)^{2}}+O\left(\mathrm{e}^{-\gamma
  l}\right)\;,
\\
p/{mc} & = & {2\pi\;n}/{l} \;,
%\\
% & = & \cosh\theta+O\left(\mathrm{e}^{-\mu l}\right)
\end{eqnarray*}
where $E_{0}(L)$ denotes  the ground state energy up to exponential finite
size corrections. As a side note, we remark that these corrections are due to vacuum polarization and
particle self-energy corrections induced by finite volume \cite{Luscher:1985dn}. 
% \footnote{These corrections are due to vacuum polarization and
%   particle self-energy 
% corrections induced by finite volume \cite{Luscher:1985dn}. } 
The dispersion above indeed describes the numerically obtained
finite size spectrum of low energy quasiparticles,  
as demonstrated in fig.~\ref{fig:DMRGopt}.
High energy deviations are mainly cut-off effects due to the
fact that the DMRG data are not close enough to the fixed point.

\begin{figure}
\begin{centering}
\includegraphics[scale=0.4,clip=true]{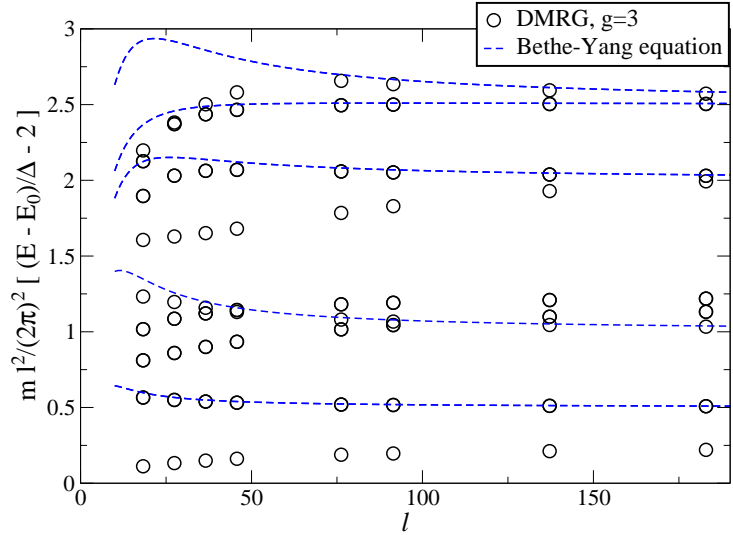}
\par\end{centering}

\caption{\label{fig:DMRGtpt}\CO 
Comparison of  rescaled DMRG data (for $g=3$) to
the two-particle levels predicted by the bootstrap $S$ matrix.
While the singlet spectrum is perfectly reproduced without 
further fitting parameter, the triplet sector cannot be fitted.}
\end{figure}

Scaling field theory predicts that
(neutral) two-particle states in the paramagnetic phase
are described
by the Bethe-Yang quantization conditions
\be
\mathrm{e}^{il\sinh\beta_{1}}S_{A\bar{A}}(\beta_{1}-\beta_{2})  =  1\;,\quad
\mathrm{e}^{il\sinh\beta_{2}}S_{A\bar{A}}(\beta_{2}-\beta_{1})  = 1\;,
\nonumber
\ee 
or, in logarithmic form,
\begin{eqnarray*}
l\sinh\beta_{1} + 2\delta_{A\bar{A}}(\beta_{1}-\beta_{2}) & = & 2\pi n_{1}\;,\\
l\sinh\beta_{2} +2\delta_{A\bar{A}}(\beta_{2}-\beta_{1}) & = & 2\pi n_{2}\;,
\end{eqnarray*}
with $n_{1}$ and $n_{2}$ integer quantum numbers, and the phase-shift
function defined as 
\begin{equation*}
\delta_{A\bar{A}}(\beta)=-\frac{i}{2} \ln S_{A\bar{A}}(\beta)\;.
\end{equation*}
The Bethe-Yang equations are nothing else than the conditions 
(\ref{eqn:paramagn_quantcond}) stated in terms of 
the notations of the scaling field theory.
The energy relative to the ground state can be computed as 
\begin{equation*}
E(L)-E_{0}(L)=\Delta(\cosh\beta_{1}+\cosh\beta_{2})+O\left(\mathrm{e}^{-\gamma' l}\right) \;,
\end{equation*}
and is accurate to all orders in $1/l$, similar to
the one-particle states.
To conform with the conventions used previously, we plot the rescaled
quantity
\begin{equation*}
\frac{l^{2}}{(2\pi)^2}\;\frac{E(L)-2 \Delta - E_{0}(L)}{\Delta}
\end{equation*}
against $l$. 
The results are shown in fig.~\ref{fig:DMRGtpt}, which
shows that the scaling field theory correctly describes the singlet
levels. In the language of the field theory, these are exactly the
two-particle levels in the $\mathcal{C}$-odd sector $\mathcal{H}_{1}$.
However, the triplet levels cannot be explained by the bootstrap $S$
matrix. In the scaling field theory, these levels are in the $\mathcal{C}$-even
sector $\mathcal{H}_{0}$ and are described by the same Bethe-Yang
equations, which means that they are exponentially degenerate with
their singlet counter-parts. While this picture is fully confirmed
by the TCSA analysis \cite{Takacs:2011xx}, it is clearly not consistent with the
DMRG spectrum.

\begin{figure}
\begin{centering}
\subfigure[ ]{
\includegraphics[scale=0.4,clip=true]{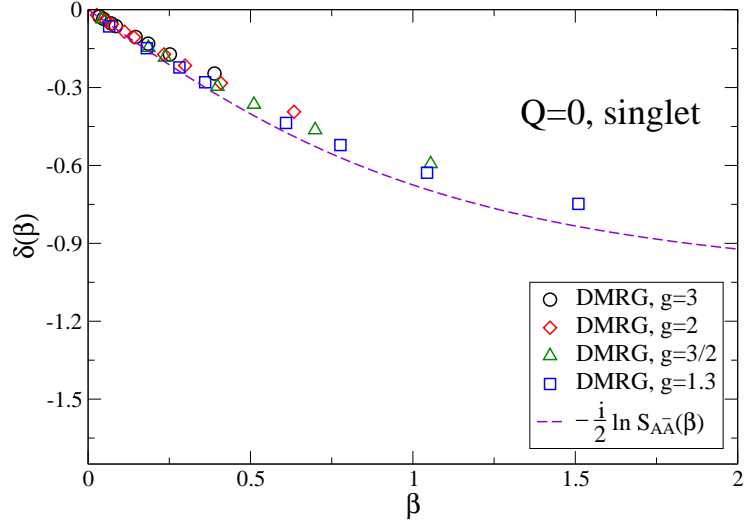}
\label{fig:delta1}
}
\ \\
\ \\
\ \\
\ \\
\subfigure[ ]{
\includegraphics[scale=0.4,clip=true]{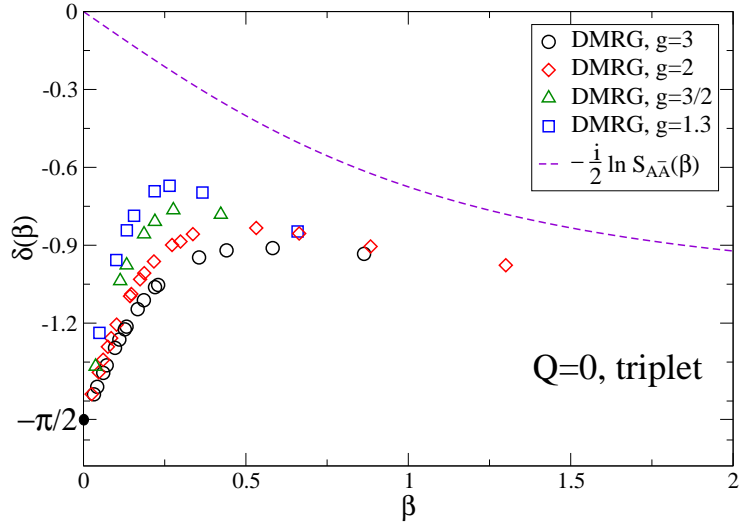}
\label{fig:delta2}
}
\par\end{centering}
\caption{\label{fig:dmrg_phaseshifts}\CO 
(a) Singlet phase-shift extracted from
DMRG and compared to $\delta_{A\bar{A}}$, as a function of rapidity
$\beta$. 
(b) Triplet phase-shift extracted from DMRG and compared to
$\delta_{A\bar{A}}$. Dashed lines are the
theoretical predictions from the bootstrap.}
\end{figure}

To see the problem more clearly, one can perform a direct comparison
of the phase-shift function to the DMRG spectrum. Provided the energy
levels $E(L)$ are known, the Bethe-Yang equations can be used
to extract the phase-shift function from them. The results are
shown in fig.~\ref{fig:dmrg_phaseshifts}. For the singlet levels
the slope of the phase-shift around the origin agrees quite well with
DMRG data, which means that the bootstrap $S$ matrix gives correctly
not only the low-energy value of the phase-shift, but also the scattering
length. For larger $\beta$ the deviations are explained by cut-off
effects since these correspond to lower values of the volume, 
closer to the scale of the lattice spacing.

However, the phase-shift extracted from the triplet states does not
agree with the bootstrap prediction at all: neither the low-energy value
nor the scattering length is consistent as figure \ref{fig:dmrg_phaseshifts} 
(b) demonstrates. Rather strikingly, however, figure 
\ref{fig:dmrg_phaseshifts} (b) shows evidence for the emergence of a new
scale: the slope of the triplet phase shift  $\delta_t(\beta)$ at
$\beta=0$ increases gradually 
as one approaches $g\to g_c=1$, which implies that, in addition to the
Compton scale $mc$, yet another small momentum scale $p^*<mc$ appears 
on the lattice. Within the most plausible scenario, this 
scale could be generated by some irrelevant operator. This would be 
indeed consistent with the fact that $p^*/mc$ appears to vanish  
as $g\to g_c$. Within this scenario, one would also naively expect  
$\delta_t(\beta)$ to scale to the scaling Potts value $\delta_{A\bar A}(\beta)$ 
for any fixed $\beta$  as $g\to g_c$. While the $\delta_t(\beta)$ curves 
indeed seem to get somewhat closer to $\delta_{A\bar A}(\beta)$ 
as $g\to g_c$, the convergence seems to be extremely slow, and the numerics 
does not yet give sufficient evidence to conclude.

\section{Conclusions}

In this work, we determined  numerically the two-particle spectrum of the $q=3$ state quantum Potts chain in
one dimension, and showed that it is in complete agreement with a simple effective field theory, 
and a corresponding asymptotic scattering matrix of an exchange form, $S(k\to 0) \to -\hat X$.
We also showed that the usual  scaling Potts field theory does not capture the observed  asymptotic behavior
and that the bootstrap S matrix reproduces  only the "singlet" part of the finite size spectrum.
This, however, does not  exclude that the  bootstrap S-matrix 
 be valid above some  momentum scale, $p^*$.
Indeed, our data support the emergence of a new small momentum scale on the lattice, 
$p^*<mc$,  and it is only below this scale that  the asymptotic (exchange) S-matrix dominates. 
This scale seems to vanish faster than the Compton scale upon approaching
 the critical point, and {\em above} this scale,  the collision 
of quasiparticles is expected to (see below) and may well be described by the scaling Potts model. 

Our numerical calculations also allowed us to check the 
duality between the  ferromagnetic and the paramagnetic states. 
We have shown that in the $Q=0$ sector the whole finite size spectrum 
obeys duality. As a consequence, the scattering lengths, 
the quasiparticle  masses and gaps also satisfy duality relations, as
demonstrated by the numerical analysis of the finite size spectra.
%The extracted scattering lengths are found to diverge at the critical  point. 

 The structure of the asymptotic S-matrix can be understood 
on simple physical grounds. For  interacting massive spinless bosons
any local interaction is relevant in one dimension~\cite{Sachdevbook}, and leads
to a scattering phase shift, $\delta=\pm\pi/2$. For particles with 
some nontrivial internal quantum number the situation is somewhat more subtle.
In the antisymmetric (singlet) channel the wave function has a
node, and therefore local interactions are asymptotically 
irrelevant, implying $\delta_s(k=0)=0$. In contrast, in the symmetric
(triplet) channel the orbital wave function does not vanish 
when the two particles approach each other. Here interactions are
relevant, and lead to $\delta_t(k=0)=\pm\pi/2$. These arguments 
immediately imply an asymptotic S-matrix of the exchange form.  
To obtain a diagonal S-matrix at $k\to 0$, one would need 
$\delta_t(k=0)=0$, which  is only possible for very special 
effective interactions between the quasiparticles~\cite{qPotts},
and is not guaranteed by  $\Z_3$ symmetry.

Though our lattice calculations are still relatively far away from the critical point itself, 
it is very hard to believe that the asymptotic theory discussed here would suddenly break down as one approaches
the critical point, $g=1$.  In this regard, we think that our results are
conclusive.   However, due to the vanishing of the gap, and the divergence of
the correlation length and the scattering lengths, our effective theory will
clearly be limited to smaller and smaller momenta, and correspondingly, to
smaller and smaller temperatures.

We should also remark that our calculations do not support the 
existence of bound states in the $\{+-\}$ and $\{-+\}$ channels: 
every state in the finite size spectrum could
be identified as an extended two-particle state. Bound states were
also absent in the paramagnetic $Q=\pm$ sectors, where  we have just observed 
extended single particle excitations at and slightly above the gap, 
$\Delta$, all in agreement with our simple effective theory. We remark that
this also agrees with the bootstrap, where particles $A$ and $\bar
A$ can be  interpreted as bound states of $AA$ or ${\bar A}{\bar A}$ 
in the spirit of Chew's ``nuclear democracy''; therefore, one does not expect
any additional state besides the multi-particle states built from $A$ and
$\bar A$ (or the corresponding kinks in the ferromagnetic phase).

We believe that the deviations between the scaling Potts field theory
and the properties of the quantum Potts chain
 are rooted at the assumption of {\em integrability}. 
Although the $q=3$ lattice Potts model is integrable at the critical point, 
$g=1$, it is believed to be {\em non}-integrable for other values of $g$. 
The scaling Potts field theory, on the other hand,  is integrable, and the bootstrap, 
together with the assumption of $\Z_3$ symmetry, leads to 
a diagonal S-matrix for $q=3$.  Similarly, in the perturbed CFT description, one perturbs the 
continuum Potts model with the leading relevant operator, 
which leaves the model integrable, and gives a spectrum
 consistent with the bootstrap S-matrix (as shown in detail by the analysis in \cite{Takacs:2011xx}).  
None of the latter methods, at least in their original form, are  
able to describe the asymptotic ($k\to 0$) properties 
of the quantum Potts chain for any fixed $g\ne 1$,
the reason  probably being that requiring  integrability conflicts with
the true non-integrable nature of the $q=3$ state  quantum Potts chain.  

We believe that to describe  the $q=3$ lattice quantum  Potts model, 
one needs to allow for perturbations or cut-off schemes which violate
integrability.  In perturbed CFT,  a possible candidate would be adding  the leading
irrelevant operator which already violates integrability. 
In this case, for any fixed finite rapidity $\beta$ 
(fixed $k/m(g)c(g)$) one expects to recover the diagonal 
S-matrix, \eqref{eq:highT_Smatrix} as $g\to 1$, and 
a new momentum scale ($p^*< mc$) 
would also be  generated, which would then separate the
regime described
by the scaling Potts model ($|k|\gg p^*$) and that described by 
the asymptotic theory  discussed here ($|k|\ll p^*$). In fact, our data seem to support this scenario, 
though unfortunately, we do not have enough numerical evidence 
to prove or disprove  convergence to the scaling Potts S-matrix for 
$p^*< |k|$. While in the singlet 
sector, shown in fig.~\ref{fig:dmrg_phaseshifts}~(a), the deviation
from the bootstrap solution is small  for intermediate rapidities $\beta
\approx 1$ and can be explained by the finite UV cutoff,  the situation 
is not so clear in the triplet case, shown in
fig.~\ref{fig:dmrg_phaseshifts}~(b), where  
the deviation from the bootstrap prediction remains considerably larger.
%than in the singlet case.
%. There, on the one hand,  
%the phase shift for a small but fixed value of $\beta \neq
%0$ seems to slowly  approach the curve corresponding to 
%the diagonal S-matrix as $g$ approaches 
%the critical point. On the other hand, while the phase shift seems
%to have reached the scaling limit for $\beta \approx 0.7$, 
%the deviation from the bootstrap prediction still remains considerably larger
%than in the singlet case. We are not entirely convinced 
%if this should be taken seriously, or if it is just due to some
%unexpected problem with our otherwise extremely accurate numerics.

The results above need also be discussed in connection to   the two-dimensional 
{\em classical} Potts model. The quantum Potts spin chain can be obtained as a
Hamiltonian limit of this system, under the assumption 
that the anisotropy introduced in the Hamiltonian ($\tau$-continuum) 
limit is irrelevant~\cite{Solyom:1986}. We remark that the anisotropy tends to infinity in 
the Hamiltonian limit as the timelike lattice spacing is taken to zero.
There is a number of results obtained by means of  the bootstrap S-matrix of the scaling
Potts theory which were compared to lattice results in the 2d classical lattice
model. In particular, universal amplitude ratios have been calculated from the
bootstrap approach \cite{Delfino:1997ag,Delfino:1999rn}, and numerical lattice
computations as well as low temperature expansions seem to agree with the
theoretical predictions~\cite{enting:2003,schur:2002_08}. Another such
quantity is the so-called static three-quark potential which also agrees with
the lattice calculations (on a triangular lattice) \cite{Caselle:2006}. 
The numerically obtained critical exponents and the central charge of the
  critical Potts spin chain also agree with the predictions of the scaling
Potts model~\cite{Hamer:1981}. However, unfortunately, these results cannot be 
conclusive regarding the structure of the S-matrix: The asymptotic S-matrix governs only a small fraction of the
energy eigenstates, and correspondingly, it is expected to have only a
very small impact on thermodynamic properties, which provide thus a very
indirect way to access the asymptotic properties of the S-matrix. 
In fact, as shown in  Ref.~\cite{schur:2002_08}, e.g.,  inclusion of the leading irrelevant
operator -- possibly responsible for the asymptotic exchange  scattering in the quantum Potts chain --
in the expansion of the susceptibility has virtually no impact on the
value of the ratio $\Gamma_L/\Gamma_T$ of the longitudinal and
transverse susceptibilities. Furthermore,  according to our results, for reduced 
temperatures $|\tau|<0.1$, one would need very large system sizes to observe 
visible signatures of the asymptotic S-matrix.

\ack
% \section{Acknowledgements}
% \vspace{0.4cm}
% {\bf Acknowledgement:} 
We would like to thank A.~Tsvelik and 
R.~Konik for helpful discussions, and G. Delfino for drawing our attention
to the numerical work concerning the 2d lattice model.  We also thank H. Saleur for 
helpful comments.  This research has been supported by Hungarian
Research Funds Nos. K73361, CNK80991, K75172, 
and Romanian grant CNCSIS PN II ID-672/2008. 
% G.Z. and \'A.R. also acknowledge support from the DFG.
We acknowledge support by the Deutsche Forschungsgemeinschaft and the Open Access
Publishing Fund of the Karlsruhe Institute of Technology.

\appendix

\section{Duality}
\label{app:duality}

We show that a one-to-one correspondence (duality relation) exits
between the energies  in the $Q=0$ subspace with PBC for 
couplings  $g \leftrightarrow 1/g$. 
To do this, we introduce two sets of basis states. The first set is
defined using the local ``spin-flip'' states 
\begin{equation}
 \vert \{ \lambda_i \} \rangle \equiv \prod_{i=1}^{L} \vert \lambda_i \rangle_i \;,
\end{equation}
where each $\lambda_i \in  \{0,1,-1\}$. Restriction to the $Q=0$
subspace implies $ \sum_{i=1}^{L} \lambda_i = 0 $. For a  given
sequence of $\{ \lambda_i \}$, there  exists another orthogonal set of
dual states, defined by using the  $\lambda_i$ as domain wall
labels, 
\begin{equation}
\vert \{ \lambda_i \} \rangle \rightarrow \widetilde  {\vert \{ \lambda_i \} \rangle_\mu} \equiv \prod_{i=1}^{L} \left\vert \mu_i=\mu + \sum_{j=1}^{i-1} \lambda_j \right\rangle_i .
\end{equation}
By construction, (since $ \sum_{i=1}^{L} \lambda_i = 0 $), 
 these states automatically satisfy PBC, 
 but they are not eigenstates of the permutation operator $\mathcal{Z}$. 
 However, we can construct states within the $Q=0$ subspace 
by defining 
\begin{equation}
\widetilde{ \vert \{ \lambda_i \} \rangle}
 \equiv \frac{1}{\sqrt{3}} [ 1 + \mathcal{Z} + \mathcal{Z}^2 ]   \vert \{ \lambda_i \} \rangle_{\mu=1}.
\end{equation}

Straightforward algebraic manipulation yields that the matrix elements
of the two terms of the Hamiltonian, $H_1=  \sum_iP^\mu_i  P^\mu_{i+1}$ 
and $H_2=  \sum_i P_i$ 
in eq.~(\ref{eq:def:H}) satisfy
the following identities: 
\begin{eqnarray}
 \widetilde {\langle \lbrace \lambda_j  \rbrace \vert}
 \sum_{i, \mu} P^\mu_i  P^\mu_{i+1}  
\widetilde {\vert  \lbrace \lambda_j' \rbrace \rangle} &=& \langle
\lbrace \lambda_j \rbrace | \sum_iP_i | \lbrace \lambda_j' \rbrace
\rangle\;,
 \nonumber 
\\
 \widetilde {  \langle \{ \lambda_j \}|} P_i  
\widetilde {|\{\lambda_j'\}\rangle} &=& \langle \lbrace \lambda_j \rbrace \vert \sum_\mu  P^\mu_i P^\mu_{i+1} \vert \lbrace \lambda_j' \rbrace \rangle. \label{eq:duality-identities}
\end{eqnarray}
Let us now assume that the states
\begin{equation}
\vert n \rangle = \sum_{\{\lambda_j\}} A^n_{\{ \lambda_j\}} \vert \{\lambda_j \} \rangle
\end{equation}
are normalized, orthogonal eigenstates 
of $H_1 + g H_2$ in the $Q=0$ subspace,
\begin{equation}
 \langle m \vert (H_1 + g H_2) \vert n \rangle = \delta_{nm} E_n(g).
\end{equation}
Then let us define the set of dual states as
\begin{equation}
\widetilde{\vert n \rangle} \equiv 
\sum_{\{\lambda_j\}} A^n_{\{ \lambda_j\}} \widetilde{\vert \{\lambda_j \} \rangle}.
\end{equation}
Using eq.~(\ref{eq:duality-identities}), we immediately 
see that that these diagonalize the dual Hamiltonian, $H_1 + 1/g H_2$,
\begin{equation}
 \widetilde{\langle m|} (H_1 + 1/g H_2) \widetilde{\vert n \rangle} = 
(1/g) \langle n \vert g H_2 + H_1 \vert m \rangle = (1/g) \delta_{nm} E_n(g)\;,
\end{equation}
yielding the duality relation, eq.~\eqref{eq:duality}.

\end{document}